\documentclass{article}
\usepackage{jalc}
\usepackage{epic,autograph,a4}

\newtheorem{Thm}{Theorem}[section]

\newtheorem{Lemma}[Thm]{Lemma}
\newtheorem{Cor}[Thm]{Corollary}

\theorembodyfont{\rm}
\newtheorem{Example}{Example}
\newtheorem{Remark}{Remark}

\newenvironment{Proof}{\rm \trivlist \item[\hskip \labelsep{\bf
Proof.}]}{\cqfd\endtrivlist}

\def\cqfd{\skip10=\parfillskip\parfillskip=0pt
\enspace\hfill\symbolecqfd\par\parfillskip=\skip10\par\medskip}
\def\symbolecqfd{\rlap{$\sqcap$}$\sqcup$}
\def\preuve{\begin{Proof}}
\def\proof{\begin{Proof}}
\def\eop{\end{Proof}}

\newenvironment{Proofof}[1]{\rm \trivlist \item[\hskip \labelsep{\bf
Proof of #1.}]}{\cqfd\endtrivlist}

\def\proofof{\begin{Proofof}}
\def\eopof{\end{Proofof}}

\let\phi=\varphi
\def\inv{^{-1}}

\def\calF{\mathcal{F}}

\def\calS{\mathcal{S}}

\font\Bb=msbm10

\def\N{\hbox{\Bb N}}

\def\Aut{\hbox{\rm Aut}}

\def\form#1{\hbox{\bf #1}}

\def\mdec{\form{mdec}}
\def\repr{\form{repr}}
\def\dist{\form{dist}}

\def\mod{\mathbf{mod}\;}

\title{On the logical definability \\
of certain graph and poset languages}

\author{Pascal Weil\,\footnotemark}

\address{LaBRI, Universit\'e Bordeaux-1 and CNRS \\
351 cours de la Lib\'eration, 33405 Talence Cedex, France}
\email{pascal.weil@labri.fr}

\abstract{We show that it is equivalent, for certain sets of finite graphs,
    to be definable in $CMS$ (counting monadic second-order logic, a
    natural extension of monadic second-order logic), and to be
    recognizable in an algebraic framework induced by the notion of
    modular decomposition of a finite graph.
    
    More precisely, we consider the set $\calF_{\infty}$ of
    composition operations on graphs which occur in the modular
    decomposition of finite graphs.  If $\calF$ is a subset of
    $\calF_{\infty}$, we say that a graph is an $\calF$-graph if it
    can be decomposed using only operations in $\calF$.  A set of
    $\calF$-graphs is recognizable if it is a union of classes in a
    finite-index equivalence relation which is preserved by the
    operations in $\calF$.  We show that if $\calF$ is finite and its
    elements enjoy only a limited amount of commutativity --- a
    property which we call weak rigidity, then recognizability is
    equivalent to $CMS$-definability.  This requirement is weak enough
    to be satisfied whenever all $\calF$-graphs are posets, that is,
    transitive dags.  In particular, our result generalizes Kuske's
    recent result on series-parallel poset languages.}
    

\begin{document}

\maketitle

\footnotetext[1]{Partial support from the following sources is
gratefully acknowledged: Project 2102-1 of IFCPAR/CEFIPRA; and AS
\textit{Automates, mod\`eles distribu\'es et temporis\'es} of
D\'epartement STIC, CNRS.}

The connection between recognizability and definability is one of the
cornerstones of theoretical computer science, going back to B\"uchi's
celebrated theorem on finite and infinite words in the 1960s (see
\cite{WT-HB}).  This theorem states the equivalence between two
fundamental properties of a language:
\begin{itemize}
    \item to be definable in monadic second order logic ($MS$),
    \item to be recognizable.
\end{itemize}
In B\"uchi's work, recognizability is defined by means of a finite
state automaton.  It is well-known that recognizability by such an
automaton is equivalent to \textit{algebraic recognizability}, that
is, to being the union of classes in a finite-index congruence.  (This is
well-known for languages of finite words but it also holds, with the
appropriate definitions, for languages of infinite words
\cite{PerrinPin95,PerrinPin03}).

One direction in which this result has been extended, is the
consideration of sets of other combinatorial structures than words. 
For instance, $MS$-definability and recognizability are equivalent for
trace languages (see \cite{Ebinger}).  Traces can be viewed as finite
posets, labeled by letters in a given alphabet $A$ --- and as such as
generalizations of words, which are $A$-labeled linearly ordered
finite sets.

A further generalization is the consideration of finite graphs.  Graph
languages have been widely studied for the description of complex
structures or behaviors; among them, languages of partially ordered
sets (\textit{posets}) are used for modeling certain aspects of
concurrency.

To handle graph languages, several definitions of algebraic
recognizability can be adopted, depending on which operations on
graphs (or constructors) are put forward (see Section~\ref{sec:rec}). 
In this paper, we consider the operations on graphs given by the
theory of the modular decomposition of graphs, which we call the
\textit{modular signature} (see Section~\ref{sec:composition}). 
Courcelle \cite{BCX} already considered this algebraic framework, and
proved that $CMS$-definability implies recognizability
(Theorem~\ref{BCX:611} below; see Section~\ref{sec:logics} about $CMS$
\textit{vs.} $MS$).  Lodaya, Weil \cite{LW-TCS00,LW-IC01} and Kuske
\cite{Kuske02} also considered a restriction of this algebraic
framework: to a particular class of graphs, the series-parallel
posets, and to the modular operations which suffice to generate them,
namely the sequential and the parallel product.  In particular, Kuske
proved \cite{Kuske02} the equivalence of $CMS$-definability and
recognizability in that case.  Our main result establishes the same
equivalence in a wider setting: it holds for a language of finite
$A$-labeled graphs $L$ provided the graphs in $L$ can be generated
from one-vertex graphs using a finite number of operations in the
modular signature; and provided that, apart from the parallel product
(or disjoint union) of graphs, these operations enjoy only a limited
amount of commutativity --- a property which we call \textit{weak
rigidity} (Section~\ref{sec:weak rigidity}).  As it turns out, this
requirement is weak enough and it is satisfied by all the finite
subsets of the modular signature which generate only posets or even
dags.  In other words, our result applies to languages of finite
$A$-labeled posets under a natural finite generation hypothesis, thus
generalizing Kuske's result.

The proof, which generalizes ideas from proofs of Kuske \cite{Kuske02}
and Courcelle \cite{BCX}, relies on the consideration of certain
tree-like normal forms for graphs (relative to the modular signature)
and uses crucially the notion of an $MS$-definable
transduction~\cite{BC97}.

Other operations have been considered, to make the set of finite
graphs into a (multi-sorted) algebra.  Among the most important such
signatures, we mention the $VR$ and the $HR$ signatures, see
\cite{BC97}.  In a series of papers (\textit{e.g.}
\cite{BCI,BCV,BC94,BCX,BC97}), Courcelle and co-authors have studied
the connection between definability and recognizability with respect
to these signatures.  A remarkable result in that direction is the
equivalence between $CMS$-definability (see Section~\ref{sec:logics}
about $CMS$ \textit{vs.} $MS$) and algebraic recognizability with
respect to $HR$, for languages of graphs of tree-width bounded by an
integer $k$ (Courcelle~\cite{BCV} for $k=2$, Kaller \cite{Kaller} for
$k=3$ and Lapoire~\cite{lapoire} for the general case).  This result
is, however, incomparable with ours.

\section{Terminology and notation}

In this section, we fix the notation and definitions which we will
use, concerning graphs or posets and the logical apparatus to specify
their properties.

\subsection{Graphs and posets}

In this paper, all graphs are assumed to be finite.

We consider directed $A$-(vertex-)labeled graphs of the form
$G=(V,E,\lambda)$ where $V$ is the (finite) set of \textit{vertices},
$E\subseteq V\times V$ is the \textit{edge relation} and the
\textit{labeling function} $\lambda\colon V\rightarrow A$ is a mapping
into a fixed alphabet $A$ (a finite, non-empty set).

When the labeling is irrelevant, we omit $\lambda$ in the description
of $G$.  Undirected graphs are considered as a special case of
directed graphs, where the edge relation is symmetric:
$(x,y)\in E$ if and only if $(y,x)\in E$.

We always assume that our graphs do not have self-loops (edges of the
form $(x,x)$), that is, $E$ is an anti-reflexive relation.  If
necessary, the presence of a self-loop at vertex $x$ can be encoded in
the letter labeling $x$.

At times, we view graphs up to isomorphism, and at other times, we
insist on so-called concrete graphs.  More precisely, when we consider
graph languages, the graphs in question are up to isomorphism.  When
we use graphs as syntactic tools to define algebraic operations (as
in Section~\ref{modular}), then the name of vertices is important,
that is, distinct graphs may well be isomorphic.

A \textit{dag}, or \textit{directed acyclic graph} is a directed graph
$G = (V,E)$ in which no path is a loop.  The graph $G$ is said to be
\textit{transitive} if the edge relation $E$ is transitive.  In
particular, $G = (V,E)$ is a transitive dag if and only if $E$ is a
partial order relation on $V$ (\textit{minus} the reflexivity part of
the relation, that is, the pairs $(x,x)$, $x\in V$, since we are
considering graphs without self-loops).  When we talk of posets, we
always refer to the associated transitive dags, so that a poset
language is a special kind of graph language.

\subsection{Logics}\label{sec:logics}

In this paper, we need to discuss logical properties of various kinds
of relational structures, beyond labeled graphs and posets as defined
above.  For this purpose, we use the classical notions as in, say,
\cite{EFT}.

In general, let $\calS$ be a finite relational signature, that is, a
finite set equipped with a mapping $\alpha\colon \calS\rightarrow \N$
into the non-negative integers, called the \textit{arity function}. 
An \textit{$\calS$-structure} is a set $X$ (the \textit{domain set} of
the structure) equipped, for each $s\in\calS$, with a relation $s^{X}$
of arity $\alpha(s)$.

When $\calS$ is fixed, logical formulas can be built using the usual 
connectives and quantifiers, and the elements of $\calS$ as 
predicates (respecting the arity function $\alpha$).

\begin{Example}
    When we discuss $A$-labeled graphs in this paper, the signature 
    consists of the binary relation $E$ (edge relation) and, for each 
    letter $a\in A$, of a unary relation $\Lambda_{a}$. The graph is 
    then, in effect, viewed as a structure with domain set its vertex 
    set $V$.
\end{Example}

When quantification is allowed only on elements of the domain, we talk
of \textit{first-order} or $FO$-formulas.  If we quantify also on sets
of elements (unary relations on the domain), we talk of
\textit{monadic second order}, or $MS$-formulas.

We also make intensive use of the following extension of monadic
second order logic.  The formalism of $MS$-formulas is enriched with
special quantifiers of the form $\exists^{\mod q}x$, where $q\ge 2$ is
an integer and $x$ is a first-order variable.  A formula of the form
$\exists^{\mod q}x\ \phi(x)$ is interpreted to mean that the
cardinality of the set of values of $x$ such that $\phi(x)$ holds is 0
mod $q$.  The resulting logic is called $CMS$ (\textit{counting
monadic second order} logic) \cite{BCI}.

It is well-known that $CMS$ is strictly more expressive than $MS$: no
$MS$ formula can express the fact that an $\calS$-structure has even
cardinality \cite{BCI}.  On the other hand, $CMS$ is designed
precisely to express this type of property.

\section{Recognizability: the algebraic framework}\label{sec:rec}

The notion of recognizability was established in the 1960s by Mezei
and Wright \cite{MW67}.  It makes sense with respect to a given
algebraic framework, that is, in a given algebra, for a given
signature.

More precisely, let $\calF$ be a signature (finite or infinite), that
is, a set equipped with a mapping $\alpha\colon \calF\rightarrow \N$
into the non-negative integers, called the \textit{arity function}. 
An \textit{$\calF$-algebra} is a set $X$ equipped, for each
$f\in\calF$, with an $\alpha(f)$-ary operation $f^{X}$.  Morphisms of
$\calF$-algebras are defined in the usual way, see
\cite{BurrisSankappanavar}.  A subset $L$ of an $\calF$-algebra $X$ is
said to be \textit{$\calF$-recognizable} (\textit{recognizable} if
there is no ambiguity) if there exists a morphism of $\calF$-algebras
$\phi$ from $X$ into a finite $\calF$-algebra such that $L =
\phi\inv(\phi(L))$.  Thus, the notion of a recognizable subset of $X$
depends on the algebraic structure considered on $X$.

There are several natural ways to view the set of all (finite) graphs
as an algebra, and hence several different notions of recognizability
(see for instance Courcelle in \cite{BC94,BC97}).  In this paper, we
operate in the algebraic framework provided by the existence and the
uniqueness of the so-called \textit{modular decomposition} of finite
graphs.  The relevant definitions are given in the next sections.

\subsection{Composition of graphs: the modular 
signature}\label{sec:composition}

If $n\ge 1$ is an integer, we denote by $[n]$ the set $\{1,\ldots,n\}$.

With each $n$-vertex graph $H$, we associate an $n$-ary operation on
graphs.  In order to properly define our algebraic setting, operations
must have a linearly ordered set of arguments, and hence we need to
view $H$ as a concrete graph with vertex set $[n]$.  In particular,
distinct isomorphic graph structures on $[n]$ define different
operations.

Let $H = ([n],F)$ and let $G_{1},\ldots,G_{n}$ be graphs.  The graph
$H\langle G_{1},\ldots,G_{n}\rangle$ is obtained by taking the
disjoint union of the graphs $G_{1},\ldots,G_{n}$, and by adding, for
each edge $(i,j)\in F$, an edge from every vertex of $G_{i}$ to every
vertex of $G_{j}$.  In other words, if $G_{i} = (V_{i},E_{i})$ for
$i=1,\ldots,n$, then $H\langle G_{1},\ldots,G_{n}\rangle = (V,E)$
where
\begin{eqnarray*}
    V &=& V_{1} \sqcup \cdots \sqcup V_{n} \cr
    E &=& E_{1} \sqcup \cdots \sqcup E_{n} \sqcup \bigsqcup_{(i,j)\in 
    F}V_{i}\times V_{j}
\end{eqnarray*}

The following 2-vertex graphs provide particularly important examples 
of such operations.

\begin{center}
\begin{picture}(75,8)(0,0)
\setvertexdiam{5}
\letvertex X1=(0,6) \drawvertex(X1){$1$}
\letvertex X2=(15,6)  \drawvertex(X2){$2$}
\letvertex X=(8,0)  \drawvertex(X){$H_{\oplus}$}
\letvertex Y1=(30,6) \drawvertex(Y1){$1$}
\letvertex Y2=(45,6)  \drawvertex(Y2){$2$}
\letvertex Y=(38,0)  \drawvertex(Y){$H_{\bullet}$}
\letvertex Z1=(60,6) \drawvertex(Z1){$1$}
\letvertex Z2=(75,6)  \drawvertex(Z2){$2$}
\letvertex ZZ1=(60,7)
\letvertex ZZ2=(75,7)
\letvertex ZZZ1=(60,5)
\letvertex ZZZ2=(75,5)
\letvertex Z=(68,0)  \drawvertex(Z){$H_{\otimes}$}
\drawedge(Y1,Y2){}
\drawedge(ZZ1,ZZ2){}
\drawedge(ZZZ2,ZZZ1){}
\end{picture}
\end{center}
The binary operation defined by $H_{\oplus}$, written $G_{1}\oplus 
G_{2}$, is simply the disjoint union of $G_{1}$ and $G_{2}$; it is 
sometimes called the \textit{parallel product} of graphs.

The binary operation defined by $H_{\bullet}$, written $G_{1}\bullet 
G_{2}$, is called the \textit{sequential product}, and it consists of 
adding to $G_{1}\oplus G_{2}$ every edge from a vertex of $G_{1}$ to a 
vertex of $G_{2}$.

The binary operation defined by $H_{\otimes}$, written $G_{1}\otimes 
G_{2}$, is called the \textit{clique product}, and it consists of 
adding to $G_{1}\oplus G_{2}$ every edge from a vertex of $G_{1}$ to a 
vertex of $G_{2}$ and every edge from a vertex of $G_{2}$ to a 
vertex of $G_{1}$.

It is immediately seen that these three operations are associative, 
and that the operations $\oplus$ and $\otimes$ are commutative.

We also note the following \textit{compositionality property}: if the
graph $H$ itself can be written as a composition, say, $H = K\langle
L_{1},\ldots,L_{r}\rangle$, then the composition $H\langle
G_{1},\ldots,G_{n}\rangle = K\langle L'_{1},\ldots, L'_{r}\rangle$ and
$L'_{j} = L_{j}\langle G_{i_{j,1}},\ldots,G_{i_{j,r_{j}}}\rangle$
where $i_{j,1},\ldots,i_{j,r_{j}}$ are the vertices of $H$ in $L_{j}$.

If $H$ cannot be written as a composition, we say that $H$ is
\textit{prime}: the compositionality property above implies that every
composition operation can be expressed in terms of operations defined
by prime graphs.

Finally we note the following \textit{commutation properties}: if $H =
([n],F)$ and $H' = ([n],F')$ are isomorphic graphs, then the
corresponding composition operations differ only by the order of the
arguments.  More precisely, if $\sigma$ is a permutation of $[n]$
which induces an isomorphism from $H'$ into $H$, then

$$H\langle G_{1},\ldots,G_{n}\rangle = H'\langle
G_{\sigma(1)},\ldots,G_{\sigma(n)}\rangle\eqno{(CP)}$$

\begin{Example}\label{ex W}
    Let $H$ and $H'$ be the following concrete (prime) graphs:
    
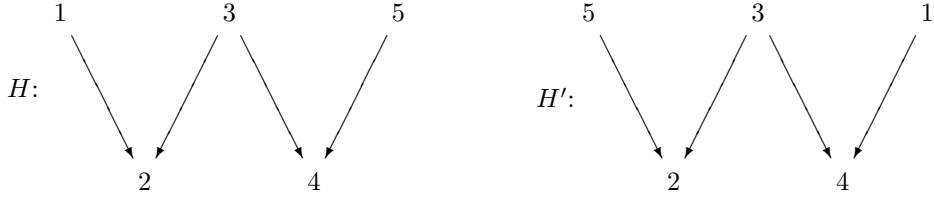
\begin{center}
\begin{picture}(82,17)(0,0)
\setvertexdiam{5}
\put(-5,8){$H$:}
\letvertex X=(0,16) \drawvertex(X){$1$}
\letvertex Z=(8,0)  \drawvertex(Z){$2$}
\letvertex Y=(16,16) \drawvertex(Y){$3$}
\letvertex T=(24,0)  \drawvertex(T){$4$}
\letvertex U=(32,16)  \drawvertex(U){$5$}
\drawedge(X,Z){}
\drawedge(Y,Z){}
\drawedge(Y,T){}
\drawedge(U,T){}
\put(45,7){$H'$:}
\letvertex X1=(50,16) \drawvertex(X1){$5$}
\letvertex Z1=(58,0)  \drawvertex(Z1){$2$}
\letvertex Y1=(66,16) \drawvertex(Y1){$3$}
\letvertex T1=(74,0)  \drawvertex(T1){$4$}
\letvertex U1=(82,16)  \drawvertex(U1){$1$}
\drawedge(X1,Z1){}
\drawedge(Y1,Z1){}
\drawedge(Y1,T1){}
\drawedge(U1,T1){}
\end{picture}
\end{center}
The permutation $(1\ 5)$ defines an isomorphism from $H'$ to $H$ 
and we have $H\langle G_{1},G_{2},G_{3},G_{4},G_{5}\rangle = H'\langle 
G_{5},G_{2},G_{3},G_{4},G_{1}\rangle$.

Similarly, the permutation $(1\ 5)(2\ 4)$ defines an automorphism of
$H$ and we have $H\langle G_{1},G_{2},G_{3},G_{4},G_{5}\rangle =
H\langle G_{5},G_{4},G_{3},G_{2},G_{1}\rangle$.
\end{Example}

In particular, we may restrict the set of concrete prime graphs
defining composition operations to having at most one representative
of every isomorphism class: in the above example, every $H'$-product
can be expressed as an $H$-product.  However, it remains necessary to
retain a concrete presentation of $H$, in order to have a unequivocal
linear order on the arguments of the corresponding operation.  Note
that this restriction does not eliminate the commutation properties
(CP): each automorphism of a prime graph induces one.

In the sequel, we select a set $\calF_{\infty}$ (the \textit{modular
signature}) consisting of the binary operations $\oplus, \otimes,
\bullet$, and of the composition operations defined by a collection of
graphs containing exactly one representative of each isomorphism class
of prime graphs with at least three vertices.  We will now view the
class of finite graphs as an $\calF_{\infty}$-algebra.

It is important to observe that $\calF_{\infty}$ is infinite, since
there are infinitely many isomorphism classes of finite prime graphs. 
In fact, almost all finite graphs are prime: more precisely, their
relative frequency among $n$-vertex graphs tends to 1, see
\cite{Mohring-Radermacher}.

\subsection{Modular decomposition}\label{modular}

The idea of the modular decomposition of a graph has been rediscovered
a number of times in the context of graph theory and of other fields
using graph-theoretic representations. We refer to
\cite{Mohring-Radermacher} for a historical survey of this question,
and to \cite{MS99} for a concise presentation. In this paper, we use
the following definitions.

Let $G = (V,E)$ be a graph.  A \textit{module} in $G$ is a subset $X$
of $V$ which interacts uniformly with its complement $V\setminus X$:
more precisely, if $v\in V\setminus X$ and $E$ contains a pair $(v,x)$
with $x\in X$, then $\{v\}\times X\subseteq E$; and dually, if
$(x,v)\in E$ for some $x\in X$, the $X\times \{v\}\subseteq E$.

We say that a module $X$ is \textit{prime} if $X\ne V$ and for every
module $Y$, either $Y\subseteq X$ or $X\subseteq Y$ or $X\cap
Y=\emptyset$.  One can show that the prime modules of a prime module
$X$ of $G$ are prime modules of $G$.  In addition, if $V$ is finite,
then the maximal prime modules of $G$ form a partition of $V$. 
Let $\equiv$ be the corresponding equivalence relation on $V$ and let
$H$ be the quotient graph $H = G/\!\equiv$: its vertex set is
$V/\!\equiv$ and its edge relation is the image of $E$ in the
projection from $V\times V$ onto $(V/\!\equiv)\times(V/\!\equiv)$. 
Then one can show that $H$ is either a prime graph $H = ([n],F)$ with
$n\ge 3$, or it is the transitive closure of one of the three
following graphs (for $n\ge 2$):

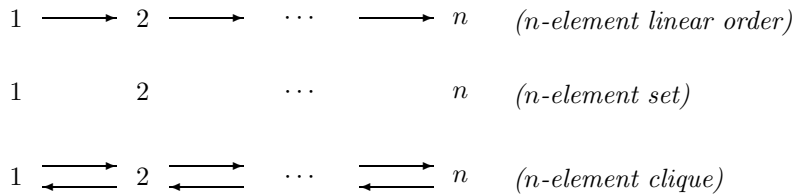
\begin{center}
\begin{picture}(75,18)(0,0)
\setvertexdiam{5}
%
%
\letvertex X1=(5,16) \drawvertex(X1){$1$}
\letvertex X2=(17,16)  \drawvertex(X2){$2$}
\letvertex X3=(29,16)
\letvertex X4=(32,16)  \drawvertex(X4){$\cdots$}
\letvertex X5=(35,16)
\letvertex X6=(47,16)  \drawvertex(X6){$n$}
\put(52,15){\it ($n$-element linear order)}
\drawedge(X1,X2){}
\drawedge(X2,X3){}
\drawedge(X5,X6){}
%
%
\letvertex Y1=(5,9) \drawvertex(Y1){$1$}
\letvertex Y2=(17,9)  \drawvertex(Y2){$2$}
\letvertex Y3=(29,9)
\letvertex Y4=(32,9)  \drawvertex(Y4){$\cdots$}
\letvertex Y5=(35,9)
\letvertex Y6=(47,9)  \drawvertex(Y6){$n$}
\put(52,8){\it ($n$-element set)}
%
%
%
\letvertex Z1=(5,1) \drawvertex(Z1){$1$}
\letvertex ZZ1=(5,2)
\letvertex ZZZ1=(5,0)
\letvertex Z2=(17,1)  \drawvertex(Z2){$2$}
\letvertex ZZ2=(17,2)
\letvertex ZZZ2=(17,0)
\letvertex ZZ3=(29,2)
\letvertex ZZZ3=(29,0)
\letvertex Z4=(32,1)  \drawvertex(Z4){$\cdots$}
\letvertex ZZ5=(35,2)
\letvertex ZZZ5=(35,0)
\letvertex Z6=(47,1)  \drawvertex(Z6){$n$}
\letvertex ZZ6=(47,2)
\letvertex ZZZ6=(47,0)
\put(52,0){\it ($n$-element clique)}
\drawedge(ZZ1,ZZ2){}
\drawedge(ZZ2,ZZ3){}
\drawedge(ZZ5,ZZ6){}
\drawedge(ZZZ2,ZZZ1){}
\drawedge(ZZZ3,ZZZ2){}
\drawedge(ZZZ6,ZZZ5){}
\end{picture}
\end{center}
In particular, if the maximal prime modules of $G$ are
$G_{1},\ldots,G_{n}$, then exactly one of the following holds:

\begin{eqnarray*}
    G &=& H\langle G_{1},\ldots, G_{n}\rangle \hbox{ for some prime
    graph $H$ with $n\ge 3$ vertices} \cr
    G &=& G_{1}\bullet G_{2}\bullet \cdots \bullet G_{n}\cr
    G &=& G_{1}\oplus G_{2}\oplus \cdots \oplus G_{n}\cr
    G &=& G_{1}\otimes G_{2}\otimes \cdots \otimes G_{n}\cr
\end{eqnarray*}    

It follows that each finite graph can be constructed
from singleton graphs, using operations from the modular signature.
Such a description of a graph is called its \textit{modular
decomposition}. In other words, the class of all finite graphs is 
an $\calF_{\infty}$-algebra generated by a single element.

Moreover, the modular decomposition of a finite graph is unique up to
the associativity of $\bullet,\oplus,\otimes$, the commutativity of
$\oplus,\otimes$, and the commutation properties (CP), based on the
non-trivial automorphisms of prime graphs. Note that the modular
decomposition of a graph can be computed in linear time
\cite{MS93,MS99,CournierHabib}.

The above discussion has been entirely concerned with unlabeled
graphs.  The generators of the $\calF_{\infty}$-algebra of $A$-labeled
graphs are simply the $A$-labeled one-vertex graphs: in other words,
the $\calF_{\infty}$-algebra of $A$-labeled graphs is generated by
$A$.

Finally, if $\calF\subseteq\calF_{\infty}$, we say that a (labeled)
graph is an \textit{$\calF$-graph} if it is in the $\calF$-algebra
generated by the singleton graphs.

\begin{Remark}\label{rk dags}
    If $\calF$ contains only dags, then the $\calF$-graphs are dags. 
    If in addition, $\calF$ consists only of transitive dags (that is,
    posets), then the $\calF$-graphs are posets.  Conversely, every
    prime graph occurring in the modular decomposition of a dag (resp. 
    a poset) is a dag (resp.  a poset).
    
    Similarly, if $\calF$ contains only undirected graphs (graphs with
    a symmetric edge relation), then the $\calF$-graphs are all
    undirected.  Conversely, every prime graph occurring in the
    modular decomposition of an undirected graph is undirected.
\end{Remark}

\subsection{Tree-like representations}\label{sec:tree-like}

We will use the following tree-like representations of an $A$-labeled
graph to account for its modular decomposition.

We first consider the tree $\mdec(G)$ (Courcelle \cite[Sec.  6]{BCX}),
whose set of nodes is the set of prime modules of $G$, and such that a
node $x$ is the parent of a node $y$ if and only if $y$ is a maximal
prime module of $x$.  Moreover, each leaf $x$ of $\mdec(G)$
(necessarily a single vertex) is labeled by $\lambda(x)\in A$, and
each inner node of $\mdec(G)$ is labeled $H$ (a prime graph in
$\calF_{\infty}$ with at least three vertices), $\bullet$, $\oplus$ or
$\otimes$, according to the fact that $x$ is an $H$-product, a
$\bullet$-product, a $\oplus$-product, or a $\otimes$-product of its
maximal prime modules.

In particular, each $\bullet$-labeled node has at least 2 children,
none of which is $\bullet$-labeled; the analogous property holds for 
each $\oplus$-labeled node and for each $\otimes$-labeled node. Each 
$H$-labeled node has $n$ children if $H$ has $n$ vertices.

In addition to this tree structure, $\mdec(G)$ also encodes the
following information.  First, there is a linear order on the children
of a $\bullet$-labeled node $x$, which comes from the modular
decomposition of $x$.  There is no such order on the children of
$\oplus$- or $\otimes$-labeled nodes.  The case of an $H$-labeled node
$x$ (where $H$ is an $n$-vertex prime graph in $\calF_{\infty}$, $n\ge
3$) is intermediary: the modular decomposition of $x$ provides an
enumeration (that is, a linear order) of the $n$ children of $x$,
which is defined up to the action of $\Aut(H)$; more formally, the
modular decomposition of $x$ provides a collection of linear orders on
the children of $x$, such that any of these order relations can be
mapped to any other one by some permutation $\sigma\in \Aut(H)$; that
is, these linear orders form an orbit under the natural action of
$\Aut(H)$.

In view of the discussion in Section~\ref{modular}, this enriched 
tree structure uniquely defines $G$.

Technically, we view $\mdec(G)$ as a relational structure whose domain
is the set of prime modules of $G$, together with the following
(interpreted) predicates:

\medskip

{\narrower\narrower\sl $\form{child}(x,y)$ if $y$ is a maximal
prime module of $x$,

$\form{label}_{a}(x)$ ($a\in A$) if $x$ is an $a$-labeled vertex of
$G$,

$\form{label}_{\oplus}(x)$ if $x$ is an $\oplus$-product of its
maximal prime modules,

$\form{label}_{\otimes}(x)$ if $x$ is an $\otimes$-product of its
maximal prime modules,

$\form{label}_{\bullet}(x)$ if $x$ is a $\bullet$-product of its
maximal prime modules,

$x < y$ if there exists a prime module $z$, with maximal prime modules
$z_{1},\ldots,z_{n}$, such that $z = z_{1}\bullet \cdots \bullet
z_{n}$, $x = z_{i}$ and $y = z_{j}$ for some $1\le i < j \le n$.

$\form{label}_{H}(x)$ (with $H\in\calF_{\infty}$ a graph with $n\ge 3$
vertices) if $x$ is an $H$-product of its maximal prime
modules,

$\form{children}_{H}(x,y_{1},\ldots,y_n)$ if $H\in\calF_{\infty}$ has
$n\ge 3$ vertices and $x = H\langle y_{1},\ldots,y_{n}\rangle$.
\par}

\medskip

Note that if $\form{label}_{H}(x)$, $\sigma\in\Aut(H)$ and
$y_{1},\ldots,y_{n}$ are the children of $H$, then
$\form{children}_{H}(x,y_{1},\ldots,y_n)$ if and only if
$\form{children}_{H}(x,y_{\sigma(1)},\ldots,y_{\sigma(n)})$.

As in Courcelle \cite{BCX} and Kuske \cite{Kuske02}, we also use the
following representation, written $\mdec'(G)$, built from $\mdec(G)$
by adding internal nodes in such a way that every $\bullet$-labeled
node has exactly two children, the first of which is not
$\bullet$-labeled.  More precisely, for each $\bullet$-labeled node
$u$ of $\mdec(G)$ with children $v_{1} < \ldots < v_{n}$ ($n\ge 3$),
we add $\bullet$-labeled nodes $u_{2},\ldots, u_{n-1}$ in such a way
that the children of $u$ are $v_{1} < u_{2}$, the children of $u_{i}$
are $v_{i} < u_{i+1}$ for $2\le i\le n-2$, and the children of
$u_{n-1}$ are $v_{n-1} < v_{n}$.  All other nodes (that is, all nodes
that are not $\bullet$-labeled) and relations are left unchanged.

The nodes of $\mdec'(G)$ can also be identified with subsets of the
vertex set of $G$, but not necessarily with prime modules of $G$. 
More precisely, with the above notation, the new vertex $u_{i}$ can be
identified with the union $\bigcup_{h=i}^n v_{h}$.

As a relational structure, $\mdec'(G)$ has domain the set of its 
nodes, and it is equipped with the following predicates, inherited 
from $\mdec(G)$:

\medskip

{\narrower\narrower\noindent\sl $\form{child}(x,y)$,
$\form{label}_{a}(x)$ ($a\in A$), $\form{label}_{\oplus}(x)$,
$\form{label}_{\otimes}(x)$, $\form{label}_{\bullet}(x)$,
$\form{label}_{H}(x)$ and $\form{children}_{H}(x,y_{1},\ldots,y_n)$ if
$H\in\calF_{\infty}$ is a graph with $n\ge 3$ vertices.  \par}

\medskip

Instead of the relation $x < y$ between distinct children of a
$\bullet$-labeled node, $\mdec'(G)$ is equipped with the binary
predicate

\medskip

{\narrower\narrower $\form{first-child}(x,y)$ if
$\form{label}_{\bullet}(x)$, $\form{child}(x,y)$ and $y$ is the first
(minimal, left-most) child of $x$.\par}

\medskip

We say that a labeled tree of the form $\mdec'(G)$, for some
$A$-labeled graph $G$, is an \textit{$\mdec'$-tree}.

\section{Recognizability vs. definability: known results}\label{sec:recdef}

The connection between definability and
$\calF_{\infty}$-recognizability was first studied by Courcelle
\cite{BCX}.  A slight modification of \cite[Theorem 6.11]{BCX} shows
the following

\begin{Thm}\label{BCX:611} Let $\calF$ be a finite subset of
$\calF_{\infty}$ and let $L$ be a language of $A$-labeled
$\calF$-graphs.  Then the following are equivalent:
\begin{itemize}
    \item $L$ is $\calF_{\infty}$-recognizable;
    \item $L$ is $\calF$-recognizable;
    \item the tree language $\mdec'(L)$ is $CMS$-definable in the
    class of $\mdec'$-trees.
\end{itemize}
Moreover, if $L$ is $CMS$-definable (in the class of graphs), then $L$
is $\calF$-recog\-nizable.
\end{Thm}

\begin{Remark}\label{MSlin}
    Courcelle's result \cite[Theorem 6.11]{BCX} is actually more
    precise: it also proves the equivalence between
    $\calF$-recog\-nizability and definability in a certain extension
    of $CMS$-logic, called $MS_{lin}$, which is well-adapted to this
    situation but lacks the good algorithmic properties of $MS$ and
    $CMS$ logic.  For our purpose, we do not need to get into the
    definition of $MS_{lin}$, and it suffices to know that
    $CMS$-definability implies $MS_{lin}$-definability.
    
    Another difference between the above statement and Courcelle's
    result is that the latter is given for unlabeled graphs and for
    particular values of $\calF$: namely the (finite) subset
    $\calF_{n}$ of all graphs in $\calF_{\infty}$ with at most $n$
    vertices for some $n$.  It is a routine verification that the same
    proof applies to $A$-labeled graphs and to any finite subset of
    $\calF_{\infty}$ -- which is necessarily contained in some
    $\calF_{n}$.
\end{Remark}
    
Courcelle shows the equivalence between $MS$-definability,
$CMS$-definability and $\calF$-recognizability when the tree language
$\mdec'(L)$ satisfies certain combinatorial properties \cite[Theorem
6.12]{BCX}, and especially when the out-degree of the internal nodes
of the elements of $\mdec'(L)$ is uniformly bounded.

This equivalence is also known to hold without restriction on the
shape of the trees in $\mdec'(L)$ for certain small values of $\calF$.

\paragraph{If $\calF = \{\bullet\}$} Since $H_{\bullet}$ is a poset,
the $\calF$-algebra consists of posets, and it is easily seen that
these posets are of the form $[n]$, equipped with the usual linear
order.  The $A$-generated $\calF$-algebra is then naturally identified
with the free semigroup $A^{+}$, i.e., the set of all finite words on
alphabet $A$ and the setting of classical language theory. 
Theorems~\ref{BCX:611} (together with \cite[Theorem 6.12]{BCX}, see
above) reduces to B\"uchi's theorem on the equivalence between
recognizability and $MS$- (and hence $CMS$-) definability.

\paragraph{If $\calF = \{\oplus\}$} The $A$-generated $\calF$-algebra
consists of the finite $A$-labeled discrete graphs (graphs without any
edges).  This algebra is naturally identified with $A^\oplus$, the
free commutative semigroup on $A$.  It is known (Courcelle \cite{BCI})
that, for languages of discrete graphs, recognizability is equivalent
to $CMS$-definability, and not to $MS$-definability.  In fact,
$MS$-definability allows only the description of finite or cofinite
discrete graph languages \cite{BCI}.

\paragraph{If $\calF = \{\otimes\}$} The $A$-generated $\calF$-algebra
consists of the finite $A$-labeled cliques.  As this is the dual
situation of discrete graphs (by edge-complementation), the same
results hold.

\paragraph{If $\calF = \{\oplus,\bullet\}$} The graphs $H_{\bullet}$
and $H_{\oplus}$ are posets, and hence every $\calF$-graph is a poset. 
These posets, called the \textit{series-parallel posets}, are exactly
those whose graph is $N$-free \cite{Gra81,VTL}.

The languages of $A$-labeled $\calF$-graphs, also called
\textit{series-parallel languages} or \textit{$sp$-languages}, were
studied by Lodaya and Weil \cite{LW-TCS00,LW-IC01} and by Kuske
\cite{Kuske02}.  In particular, Kuske showed that for $sp$-languages,
$\calF$-recognizability is equivalent to $CMS$-definability
\cite[Theorem 6.15]{Kuske02}.

\section{Weakly rigid signatures and $CMS$-definability}

Our main result, Theorem~\ref{new theorem} below, generalizes the
results of the previous section: it asserts the equivalence between
$\calF$-recognizability and $CMS$-definability for more general finite
subsignatures $\calF$ of the modular signature, and in particular for
every finite subsignature consisting only of dags.

\subsection{Weakly rigid signatures}\label{sec:weak rigidity}      

Let $H = ([n],F)$ be a prime graph ($n\ge 2$).  We say that $H$ is
\textit{weakly rigid} if the automorphism group $\Aut(H)$ does not act
transitively on $[n]$. That is: there are vertices $i\ne j$ of $H$ 
such that no automorphism of $H$ maps $i$ to $j$.

\begin{Example}\label{wr ops}
    For each $n\ge 2$, the directed cycle of length $n$, $C_{n}$, is
    not weakly rigid.  Indeed, every cyclic permutation of $[n]$
    defines an automorphism of $C_{n}$.  The same holds for $D_{n}$,
    the undirected cycle of length $n$ (for $n=2$ or $n\ge 5$: $D_{3}$
    and $D_{4}$ are not prime\dots).  Note that $D_{2} = H_{\otimes}$.
    
    The graph $H_{\bullet}$ is weakly rigid.  The graph $H$ from
    Example~\ref{ex W} is weakly rigid since it has a single
    non-trivial automorphism, namely $(1\ 5)(2\ 4)$.  In particular,
    no automorphism of $H$ can map vertex 1 to vertex 2.
    
    This graph $H$ is a particular case of a more general situation:
    every prime dag is weakly rigid, except for $H_{\oplus}$.  Indeed,
    in such a dag there are maximal elements (for the partial order
    relation obtained by taking the reflexive transitive closure of
    the edge relation) and not every vertex is maximal.  The weak
    rigidity follows from the simple observation that every
    automorphism of a dag preserves the maximal elements.
    
    More generally, every prime graph in which the in-degree or the
    out-degree is not uniform, is weakly rigid.
\end{Example}

We say that a subset $\calF$ of $\calF_{\infty}$ is a \textit{weakly
rigid signature} if $\calF$ is finite, $\calF$ contains at most one of
the operations $\otimes$ and $\oplus$, and every other operation in
$\calF$ is associated with a weakly rigid prime graph.

\begin{Example}\label{wr sig}
    In view of Remark \ref{rk dags} and Example \ref{wr ops}, every 
    finite subsignature of $\calF_{\infty}$ consisting only of dags is 
    weakly rigid.
\end{Example}    

We now state our main theorem.

\begin{Thm} \label{new theorem}
    Let $\calF$ be a weakly rigid signature and let $L$ be a language
    of $A$-labeled $\calF$-graphs.  Then $L$ is $CMS$-definable if and
    only if $L$ is $\calF$-recognizable.
\end{Thm}

\begin{Remark}
    Theorem~\ref{new theorem} generalizes Kuske's result on
    $sp$-languages \cite{Kuske02}, see Section~\ref{sec:recdef}.  It
    constitutes a refinement of Courcelle's theorem \cite[Theorem
    6.11]{BCX} (see Remark~\ref{MSlin}), which only asserts the
    equivalence between $\calF$-recognizability and
    $MS_{lin}$-definability.  However, Courcelle's result does not
    assume that $\calF$ is weakly rigid.
\end{Remark}

In view of the importance of poset languages, it is worth stating the
following particular case (see Example~\ref{wr sig}) of
Theorem~\ref{new theorem}.

\begin{Cor}
    Let $\calF$ be a finite subset of $\calF_{\infty}$ such that every
    $\calF$-graph is a poset (resp.  a dag).  A language of
    $A$-labeled $\calF$-posets is $\calF$-recognizable if and only if
    it is $CMS$-definable.
\end{Cor}    

\subsection{Proof of Theorem~\ref{new theorem}}               

The proof of Theorem~\ref{new theorem}, given below, uses the notion
of an $MS$-definable transduction introduced by Courcelle
\cite[Section 2]{BCV}.  The definition of these transductions is given
in Section~\ref{proof of MS-definable} together with the proof of the
following theorem.

\begin{Thm} \label{MS-definable}
    Let $\calF$ be a weakly rigid signature.  The mapping which
    assigns to each $A$-labeled $\calF$-graph $G$ the tree $\mdec'(G)$
    is $MS$-definable.
\end{Thm}

Note that Courcelle shows that if $\calF$ is any finite subset of
$\calF_{\infty}$, then the mapping which assigns to each linearly
ordered $A$-labeled $\calF$-graph $G$ the tree $\mdec'(G)$ is an
$MS$-transduction \cite[Corollary 6.9]{BCX}.  With our extra
assumption on $\calF$, we are able to dispense with the heavy
requirement of considering only linearly ordered graphs.

\begin{Proofof}{Theorem~\ref{new theorem}}
    Theorem~\ref{BCX:611} proves half of the equivalence; namely, it 
    asserts that every $CMS$-definable language of $A$-labeled 
    $\calF$-graphs is $\calF$-recognizable (without assuming that 
    $\calF$ is weakly rigid).
    
    In order to prove the converse, we assume that $L$ is an
    $\calF$-recognizable language of $A$-labeled $\calF$-graphs.  By
    Theorem~\ref{BCX:611}, $\mdec'(L)$ is $CMS$-definable in the
    language of $\mdec'$-trees.
    
    Now, the inverse image of a $CMS$-definable subset by an
    $MS$-transduction is $CMS$-definable \cite[Corollary 2.7]{BCV}. 
    Thus Theorem~\ref{new theorem} is an immediate consequence of
    Theorem~\ref{MS-definable}.
\eopof

\subsection{Proof of Theorem~\ref{MS-definable}}\label{proof of 
MS-definable}

We now fix a finite weakly rigid signature $\calF$.  For
convenience, we assume that $\otimes\not\in\calF$; the proof would be
completely similar if we assumed that $\oplus\not\in\calF$.

Let us first explain how we use the hypothesis that $\calF$ is weakly
rigid: for each (concrete) prime graph $H = ([n],F) \in \calF$ with
$n\ge 3$ vertices, we fix a proper, non-empty subset $\dist(H)$ of
$[n]$ of so-called \textit{distinguished vertices}, which is preserved
under the action of $\Aut(H)$.  Such a set exists by assumption: we
can choose an orbit of $[n]$ under the action of $\Aut(H)$, or in
the case of a dag the set of maximal vertices, etc.  With this choice
of $\dist(H)$, we define in an $\mdec'$-tree a new binary predicate
$\form{dist-child}_{H}(x,y)$, interpreted to mean that $x$ is
$H$-labeled, there exist $y_{1},\ldots,y_{n}$ such that
$\form{children}_{H}(x,y_{1},\ldots,y_{n})$, and $y = y_{i}$ for some
$i\in\dist(H)$.

Thus, if $G$ is an $\calF$-graph, each node of $\mdec'(G)$ that is
neither a leaf nor is labeled $\oplus$ has some distinguished children
and some non-distinguished ones.  By convention, the distinguished
vertex of $H_{\bullet}$ is the origin of the single edge --- so that
the distinguished child of a $\bullet$-labeled node is its first
child.

\begin{Remark}
    We have seen that the order of children of an $H$-labeled node is
    defined only up to the action of $\Aut(H)$: the notion of
    distinguished children is devised precisely to take into account
    this flexibility.  Weakly rigid operations are precisely those for
    which some children can be designated unambiguously as
    \textit{distinguished}.
\end{Remark}
    
\begin{Remark}
    Note that, the set $\dist(H)$ being fixed, $\form{dist-child}_{H}$
    is not truly a new predicate to be added in the signature of
    $\mdec'$-trees, but rather an abbreviation for a first-order
    formula in the language of $\mdec'$-trees.
    
    Since we are dealing only with finite signatures, we do not bother 
    with a formal mechanism to choose the sets $\dist(H)$. If we had to 
    work with an infinite signature, it would be important to introduce a 
    more formal definition of distinguished children, for instance 
    based on a logical formula (on $H$) describing these distinguished 
    children. For instance, in the case of dags, one could always 
    consider the maximal elements (or equivalently, the vertices of 
    in-degree zero).
\end{Remark}

Now we need to show that if $G$ is an $A$-labeled $\calF$-graph, then
$\mdec'(G)$ can be represented in $G$, its domain can be specified in
the monadic second-order logic of graphs, and the predicates of the
language of $\mdec'$-trees can be specified in the same language.

%

More precisely, following the definition in \cite[Section 2]{BCV}, we
need to verify the following (complex) condition: There exist integers
$k,n\ge 0$ and $MS$-formulas in the language of graphs $\phi(\vec X)$,
$\psi_{1}(x,\vec X),\ldots,\psi_{k}(x,\vec X)$, where $\vec
X = (X_{1},\ldots,X_{n})$ is a vector of second-order variables called
\textit{parameters};  and, for each
$\ell$-ary predicate $q$ in the language of $\mdec'$-trees and each
length $\ell$ vector $\vec{\imath}$ of integers in $[0,k]$ there
exists an $MS$-formula
$\theta_{q,\vec{\imath}}(x_{1},\ldots,x_{\ell},\vec X)$ with the
following property.

First we define, for each $A$-labeled $\calF$-graph $G$ and for each
assignment $\vec\gamma$ of values to the vector of variables $\vec X$
such that $G$ satisfies $\phi(\vec\gamma)$, the structure
$\repr_{\vec\gamma}(G)$ by letting:
    
    \begin{itemize}
	\item the domain of $\repr_{\vec\gamma}(G)$ consists of the
	pairs $(v,i)$ such that $v\in V$, $0\le i\le k$ and $G$
	satisfies $\psi_{i}(v,\vec\gamma)$;
	\item if $\vec{\imath} = (i_{1},\ldots,i_{\ell})$ is a vector
	of integers in $[0,k]$, $q$ is an $\ell$-ary predicate and
	$(v_{1},i_{1}),\ldots,(v_{\ell},i_{\ell})$ are elements of the
	domain of $\repr_{\vec\gamma}(G)$, then
	$\repr_{\vec\gamma}(G)$ satisfies
	$q((v_{1},i_{1}),\ldots,(v_{\ell},i_{\ell}))$ if and only if
	$G$ satisfies
	$\theta_{q,\vec{\imath}}(v_{1},\ldots,v_{\ell},\vec\gamma)$.
    \end{itemize}
The condition to be verified is finally that for each $G$, there exists
an assignment $\vec\gamma$ such that $\repr_{\vec\gamma}(G)$ is isomorphic
to $\mdec'(G)$.

For this purpose, we first encode the inner nodes of an $\mdec'$-tree
in its leaves. This idea was first introduced by Potthoff and Thomas
\cite{Potthoff-Thomas}, and used also in \cite[Section 5]{BCX}. In
fact, we cannot use a single encoding as in the works cited, and we
construct a collection of four such encodings as in Kuske's
\cite{Kuske02}. As it turns out, it is more convenient to define the
inverse of these encodings: this is done in Section~\ref{encoding}.

This construction allows us to consider a structure isomorphic to
$\mdec'(G)$, and defined within $G$ in the form required by the
definition of $MS$-transductions (Section~\ref{representing}).  It
then suffices to verify that the domain of $\form{repr}(G)$ and the
predicates in this structure are expressible by means of $MS$-formulas
on the graph $G$, which is done in Section~\ref{defining}.

We strongly rely on the fact that the nodes of the tree $\mdec'(G)$ are
naturally viewed as subsets of $V$ (see Section~\ref{sec:tree-like}),
and that $V$ is both the vertex set of $G$ and the set of leaves of
$\mdec'(G)$. In particular, in the encodings we construct, each inner
node is represented by a leaf of $\mdec'(G)$ and not by a pair of
leaves as in Courcelle \cite{BCX} or Kuske \cite{Kuske02}.

\subsubsection{Encoding the nodes of an $\mdec'$-tree}\label{encoding}

Let $T$ be an $\mdec'$-tree. We partition its set $N$ of nodes as 
follows: we let $N_{0}$ be the set of leaves; $N_{1}$ be the set of 
$\oplus$-labeled nodes all of whose children are leaves; $N_{2}$ be 
the set of $\oplus$-labeled nodes not in $N_{1}$; and $N_{3}$ be the 
complement of $N_{0}\cup N_{1}\cup N_{2}$. That is, $N_{3}$ consists 
of the $\bullet$-labeled and the $H$-labeled nodes, where $H\in\calF$ 
has arity at least 3; in particular, the nodes in $N_{3}$ have 
distinguished and non-distinguished children.

Next we define mappings $\nu$ (resp.  $\mu_{0}$, $\mu_{1}$, $\mu_{2}$,
$\mu_{3}$) from $N$ (resp.  $N_{0}$, $N_{1}$, $N_{2}$, $N_{3}$) to the
powerset of $N_{0}$ as follows.

\medskip

{\narrower\narrower
\sl
If $x\in N_{0}$, we let $\nu(x) = \mu_{0}(x) = \{x\}$.

If $x\in N_{1}$, we let $\nu(x) = \mu_{1}(x)$ be the set of children
of $x$.

If $x\in N_{2}$, we let $\nu(x) = \bigcup\nu(y)$ where the union runs
over the children $y$ of $x$; and we let $\mu_{2}(x) =
\bigcup\mu_{3}(y)$ where the union runs over the children of $x$ which
are not leaves, and hence which are in $N_{3}$.

If $x\in N_{3}$, we let $\nu(x) = \bigcup\nu(y)$ where the union runs
over the non-distinguished children $y$ of $x$; and we let $\mu_{3}(x)
= \bigcup\nu(y)$ where the union runs over the distinguished children
$y$ of $x$.
\par}

\medskip

It is easily verified that these mappings are well-defined and that,
for each $x\in N_{i}$ ($i=0,1,2,3$), $\nu(x)$ and $\mu_{i}(x)$ are
non-empty sets of leaves.

For each leaf $x$ of $T$ we denote by $\rho(x)$ the unique path from
$x$ to the root of $T$.  The following lemma is a simple rewriting of
the definition of $\nu$ and the $\mu_{i}$.

\begin{Lemma}
    Let $i\in\{0,1,2,3\}$, let $y\in N_{i}$ be a node of $T$ and let $x\in
    N_{0}$ be a leaf.  The following are equivalent:
    \begin{itemize}
	\item if $i=0$, then $x\in\nu(y)$ iff $x\in\mu_{0}(y)$ iff
	$x=y$;
	\item if $i=1$, then $x\in\nu(y)$ iff $x\in\mu_{1}(y)$ iff $y$
	is the parent node of $x$, $y$ is labeled $\oplus$ and all its
	children are leaves;
	\item if $i=2$ or $i=3$, then $x\in\nu(y)$ iff $y$ sits along
	$\rho(x)$ and every node in $N_{3}$ between $x$ and $y$ along
	$\rho(x)$ is reached from one of its non-distinguished
	children;
	\item if $i=2$, then $x\in\mu_{2}(y)$ iff $y$ sits along
	$\rho(x)$, $y$ is reached from one of its children in $N_{3}$,
	say $z$, $z$ is reached from one of its distinguished
	children, and every node in $N_{3}$ along $\rho(x)$ and before
	$z$ is reached from one of its non-distinguished children;
	\item if $i=3$, then $x\in\mu_{3}(y)$ iff $y$ sits along
	$\rho(x)$, $y$ is reached from one of its distinguished
	children, and every other node in $N_{3}$ along $\rho(x)$ and
	before $y$ is reached from one of its non-distinguished
	children;
	\item if $i=2$, then $x\in\mu_{2}(y)$ iff $x\in\mu_{3}(z)$ for some 
	child $z$ of  $y$ in $N_{3}$.
\end{itemize}	
\end{Lemma}

It follows from this lemma that for each $i$, $\mu_{i}$ is the inverse
image of a partial onto mapping $\kappa_{i}\colon N_{0}\rightarrow
N_{i}$.  More precisely, we have:

    \begin{itemize}
	\item $\kappa_{0}(x) = x$;
	\item $\kappa_{1}(x)$ is the parent node of $x$ --- if that
	node is labeled $\oplus$ and all its children are leaves;
	\item $\kappa_{3}(x)$ is the first node $y\in N_{3}$ along
	$\rho(x)$ (starting from the leaf $x$) reached from one of its
	distinguished children --- if there is such a node $y$;
	\item $\kappa_{2}(x)$ is the parent node of $\kappa_{3}(x)$
	--- if $\kappa_{3}(x)$ exists and its parent node is labeled
	$\oplus$.
\end{itemize}

\subsubsection{Representing an $\mdec'$-tree in its
leaves}\label{representing}

Let $T$ be an $\mdec'$-tree as above.  Let $\repr_{0}(T)$ be the
following structure, with the same signature as $\mdec'$-trees.  The
domain of $\repr_{0}(T)$ is the set
$$\{(x,i) \mid x\in N_{0},\ 0\le i\le 3,\ \kappa_{i}(x) \hbox{ is
defined}\}.$$

We let:
\begin{itemize}
    \item $\form{label}_{a}((x,i))$ if and only if $i=0$ and
    $\form{label}_{a}(x)$ in $T$ (for each $a\in A$);
    \item $\form{label}_{\oplus}((x,i))$ if $i=1$ or $i=2$;
    \item $\form{label}_{\bullet}((x,i))$ if and only if $i=3$ and
    $\form{label}_{\bullet}(\kappa_{3}(x))$ in $T$;
    \item $\form{label}_{H}((x,i))$ if and only if $i=3$ and
    $\form{label}_{H}(\kappa_{3}(x))$ in $T$ (where $H$ is a prime
    graph in $\calF$ with at least 3 vertices).
    \item $\form{child}((x,i),(y,j))$ if
    $\form{child}(\kappa_{i}(x),\kappa_{j}(y))$ in $T$;
    \item $\form{first-child}((x,i),(y,j))$ if
    $\form{first-child}(\kappa_{i}(x),\kappa_{j}(y))$ in $T$;
    \item
     $\form{children}_{H}((x,i), (y_{1},j_{1}), \ldots, (y_{r},j_{r}))$
    if $\form{children}_{H}(\kappa_{i}(x), \kappa_{j_{1}}(y_{1}),
    \ldots, \kappa_{j_{r}}(y_{r}))$;
    \item $\form{dist-child}_{H}((x,i),(y,j))$ if
    $\form{dist-child}_{H}(\kappa_{i}(x), \kappa_{j}(y))$ in $T$.
\end{itemize}

Note that the mappings $\kappa_{i}$ are usually many-to-one, so 
that $\repr_{0}(T)$ is not isomorphic to $T$ (and it is not an 
$\mdec'$-tree).

Let us say that two elements $(x,i)$ and $(y,j)$ of the domain of
$\repr_{0}(T)$ are $\equiv$-equivalent if $i=j$ and $\kappa_{i}(x) =
\kappa_{i}(y)$.  It is easily verified that if
$X_{0},\ldots,X_{3}\subseteq N_{0}$ are such that $X =
\bigcup_{i=0}^{3}(X_{i}\times\{i\})$ is a set of representatives of
the $\equiv$-classes, then the restriction of $\repr_{0}(T)$ to $X$ is
isomorphic to $T$.  This substructure of $\repr_{0}(T)$ (which depends
on the choice of the $X_{i}$, but is unique up to isomorphism), is
denoted --- abusing notation --- by $\repr(T)$.

\subsubsection{$\mdec'(G)$ is $MS$-definable}\label{defining}

We now consider the case where the tree $T$ arises from the modular
decomposition of an $A$-labeled $\calF$-graph $G = (V,E,\lambda)$, $T
= \mdec'(G)$.

Recall that the set $N_{0}$ of leaves of $T$ is equal to $V$ and that,
more generally, the nodes of $T$ are particular subsets of $V$.  In
view of their definition, the mappings $\kappa_{i}$ can be described
as follows.

\begin{Lemma}\label{English description of kappa}
    Let $x\in V$. 
    \begin{itemize}
	\item $\kappa_{0}(x) = \{x\}$.
	\item If the least disconnected prime module $P$ containing
	$x$ is discrete, then $\kappa_{1}(x) = P$; otherwise
	$\kappa_{1}(x)$ is not defined.
	\item If there exists a connected node $P$ containing $x$ such
	that $x$ lies in a distinguished child of $P$ and, for every
	non-trivial connected node $Q$ containing $x$ and properly
	contained in $P$, $x$ lies in a non-distinguished child of
	$Q$, then $\kappa_{3}(x) = P$; otherwise $\kappa_{3}(x)$ is
	not defined.
	\item If $\kappa_{3}(x)$ is defined and the least node $P$
	properly containing it is disconnected, then $\kappa_{2}(x) =
	P$; otherwise $\kappa_{2}(x)$ is not defined.
\end{itemize}
\end{Lemma}

We use the following collection of $MS$-definable properties of an
$A$-labeled $\calF$-graph $G = (V,E,\lambda)$.  Upper-case letters
$X,Y,\ldots$ represent subsets of $V$ or second-order variables, and
lower-case letters $x,y,\ldots$ represent elements of $V$ or
first-order variables. $H = ([n],F)$ is a prime graph in $\calF$ with 
$n\ge 3$.

{\narrower\narrower\parindent=0pt\parskip=3pt\small\sl
$\form{singleton}(X,x)$ if $X=\{x\}$.

$\form{label}_{a}(X)$, where $a\in A$, if $X$ is an $A$-labeled node 
of $\mdec'(G)$, that is, $X = \{x\}$ and $\lambda_{a}(x)$ for some $x$.

$\form{partition}(X,X_{1},\ldots,X_{n})$ if $(X_{1},\ldots,X_{n})$ is
a partition of $X$, that is, $X$ is the disjoint union of the $X_{i}$
and each $X_{i}$ is non-empty.  {\rm (To be completely correct, this
predicate should be replaced by $(n+1)$-ary predicates
$\form{partition}_{n}$, for $n=2$ and for each $n$ such that an
$n$-vertex graph lies in $\calF$; furthermore, everyone of these
predicates can be expressed in terms of $\form{partition}_{2}$.)}

$\form{module}(X)$ if $X$ is a module of $G$.  This is equivalent to
\begin{eqnarray*}
\forall y\not\in X 
& & \exists x\in X\ E(x,y) \Rightarrow \forall x\in X\ E(x,y) \cr
& \land &
\exists x\in X\ E(y,x) \Rightarrow \forall x\in X\ E(y,x).
\end{eqnarray*}

$\form{pmodule}(X)$ if $X$ is a prime module of $G$, that is,
$$\form{module}(X)\land(\forall Y\
\form{module}(Y)\Longrightarrow(Y\subseteq X)\lor(X\subseteq
Y)\lor(X\cap Y=\emptyset)).$$

$\form{disconnected}(X)$ if $X$ is disconnected, that is,
$$\exists Y\ \exists Z\  
\form{partition}(X,Y,Z)\land (\forall y\in Y\ \forall 
z\in Z\ \neg E(y,z) \land \neg E(z,y)).$$

$\form{connected}(X)$ if $X$ is connected, that is
$\neg\form{disconnected}(X)$.

$\form{label}_{\oplus}(X)$ if $X$ is an $\oplus$-labeled node of 
$\mdec'(G)$, that is, a disconnected prime module.

$\form{children}_{H}(X,X_{1},\ldots,X_{n})$ if $X, X_{1},\ldots,
X_{n}$ are prime modules and $X = H\langle X_{1},\ldots,X_{n}\rangle$. 
The latter assertion is equivalent to
\begin{eqnarray*}
    & \form{partition}&(X,X_{1},\ldots,X_{n}) \cr
    & \land & \bigwedge_{(i,j)\in F}\forall x_{i}\in X_{i}\
    \forall x_{j}\in X_{j}\ E(x_{i},x_{j}) \cr
    & \land & \bigwedge_{(i,j)\not\in F}\forall x_{i}\in X_{i}\
    \forall x_{j}\in X_{j}\ \neg E(x_{i},x_{j})
\end{eqnarray*}


$\form{label}_{H}(X)$ if $X$ is an $H$-labeled node of $\mdec'(G)$,
that is,
$$\exists X_{1}\ \ldots\ \exists X_{n}\ 
\form{children}_{H}(X,X_{1},\ldots,X_{n}).$$

$\form{suffix}(X,Z)$ if $Z$ is a non-trivial (sequential) suffix of
$X$, that is,
$$\exists Y\ \form{partition}(X,Y,Z) \land (\forall y\in Y\ \forall
z\in Z\ E(y,z) \land \neg E(z,y)).$$

$\form{sequential}(X)$ if $X$ is a sequential product, that is, it has 
a non-trivial suffix.

$\form{initial}(X,Y)$ if $Y$ is the first (least) prefix of $X$, that 
is, $X\setminus Y$ is a suffix of $X$ and $Y$ itself is not sequential.

$\form{label}_{\bullet}(X)$ if $X$ is a $\bullet$-labeled node of
$\mdec'(G)$, that is, $X$ is sequential and either it is prime module, 
or it is a suffix of a sequential prime module.

$\form{node}(X)$ if $X$ is a node of $\mdec'(X)$, that is, $X$ is
either a singleton, or a $\bullet$-labeled node, or an
$\oplus$-labeled node, or an $H$-labeled node for some prime graph
$H\in \calF$ with at least three vertices.

$\form{child}^{*}(X,Y)$ if $X$ and $Y$ are nodes and $X$ is an 
ancestor of $Y$ in $\mdec'(X)$, that is, $Y$ is properly contained in 
$X$.

$\form{child}(X,Y)$ if $X$ and $Y$ are nodes and $X$ is a (the) 
minimal ancestor of $Y$.

$\form{dist-child}(X,Y)$ if $Y$ is a distinguished child of $X$, that
is $\form{child}(X,Y)$ and, either $X$ is $\bullet$-labeled and $Y$ is
the first prefix of $X$, or $X$ is $H$-labeled (for some $H\in \calF$
with at least three vertices) and
$$\exists X_{1}\ \ldots\ \exists X_{n}\ 
\form{children}_{H}(X,X_{1},\ldots,X_{n}) \land
\bigvee_{i\in \dist(H)}Y=X_{i}.$$ 
}

Together with Lemma~\ref{English description of kappa}, this list of
definable properties shows that the formulas $X = \kappa_{0}(x)$, $X =
\kappa_{1}(x)$, $X = \kappa_{2}(x)$ and $X = \kappa_{3}(x)$ can be
expressed in monadic second-order formulas.

We now verify formally that the mapping $G\mapsto\mdec'(G)$ is
$MS$-definable.  As established in Section~\ref{representing}, it
suffices to study the mapping $G\mapsto\repr(T)$ where $T= \mdec'(G)$. 
With reference to the definition given at the beginning of
Section~\ref{proof of MS-definable}, we let $k = 3$ and $n=4$, that
is, the definition makes use of 4 parameter second-order variables
$X_{0},\ldots,X_{3}$, which will stand for sets of representatives of
the $\equiv$-classes among the domain elements of the form $(x,0),
\ldots, (x,3)$.

Since $\repr(T)$ is defined for every $G$, the role of formula
$\phi(X_{0},\ldots,X_{3})$ is solely to make sure that the assignment
of values to the parameter variables is correct.  It is chosen to
express, for $i=0,1,2,3$, that $X_{i}$ is contained in the domain of
$\kappa_{i}$, it does not contain distinct elements with the same
$\kappa_{i}$-image, and for each $x$, if $\kappa_{i}(x)$ is defined
then there exists $y\in X_{i}$ such that
$\kappa_{i}(x)=\kappa_{i}(y)$.

For $i\in[0,3]$, the formula $\psi_{i}(x,\vec X)$ is $\psi_{i} = (x\in
X_{i})$.

Finally, each relation $q$ of arity $r$ in the description of
$\repr(T)$ is as in $\repr_{0}(T)$, and it can be $MS$-defined using
the list of properties given above.

This concludes the proof of Theorem~\ref{MS-definable}.

\subsection*{Conclusions}

We have proved the equivalence between $\calF$-recognizability and
$CMS$-defin\-ability for a large class of finite subsignatures $\calF$
of the modular signature $\calF_{\infty}$.  We have not however proved
that this equivalence does not hold for the other subsignatures!  In
fact, Courcelle conjectured that $CMS$-definability is strictly weaker
than $MS_{lin}$-definability for general graphs \cite[Conjecture
7.3]{BCX}.  One closely related, yet stronger question is to find out
whether there exists a finite subset $\calF\subseteq\calF_{\infty}$
such that, for sets of $\calF$-graphs, $CMS$-definability is strictly
weaker than $\calF$-recognizability (or than
$\calF_{\infty}$-recognizability, see Theorem~\ref{BCX:611}). 
Theorem~\ref{new theorem} does not solve this problem, it only
designates a large class of finite signatures for which the two
notions are equivalent.  As pointed out by Courcelle, an archetypal
setting to discuss this conjecture is given by cographs, that is, the
$\calF$-graphs for $\calF = \{\otimes,\oplus\}$.

One can also investigate which natural $\calF$-recognizable classes of
$\calF$-graphs are characterized by algebraic properties of the finite
$\calF$-algebras recognizing them.  This type of investigation is
highly developed in the field of word languages (see \cite{Pin96}),
but also of trace languages \cite{GuaianaRS,EbMu96,Ebinger}, infinite word
languages \cite{PerrinPin03}.  Lodaya and Weil showed, in this
fashion, that the recognizable languages of series-parallel posets of
bounded width are characterized algebraically (and in an effective
fashion) \cite{LW-TCS00}.  Kuske studied the first-order definable
languages of series-parallel posets, and gave an algebraic
characterization for them in the bounded width case \cite{Kuske02}. 
The general (arbitrary-width) case remains open.

Finally, one could ask for a model of automata to handle
$\calF$-recognizable sets of $\calF$-graphs. Let us mention that
\cite{LW-IC01} proposes a model of automata which can be used to
process $\calF$-graphs if $\calF$ contains neither $\oplus$ nor
$\otimes$. To be precise, the input for these automata is an
$\mdec$-tree or an $\mdec'$-tree, but this distinction is not
algorithmically crucial if we remember that such a tree can be computed
in linear time from the graph itself (see
\cite{MS93,MS99,CournierHabib}). The accepting power of these automata
matches exactly that of $\calF$-recognizability. On the other hand, if
$\oplus\in\calF$, then the same automaton model can be used but it is
strictly more powerful than $\calF$-recognizability. Eliminating in
this way the use of an associative commutative operation reduces the
interest of the construction, and the question remains open to propose
a different automaton model for $\calF$-graphs in general --- or for
series-parallel posets in particular, that is for the situation where
$\calF = \{\bullet,\oplus\}$, studied especially in
\cite{LW-TCS00,LW-IC01}.


\end{document}